\documentclass[12pt,letterpaper]{article}
\usepackage{amsmath,amssymb}
\usepackage{graphicx,color}
\usepackage{hyperref}

\usepackage[bf]{caption}

\newcommand{\rr}{\mathbb{R}}
\newcommand{\zz}{\mathbb{Z}}

\newcommand{\be}{\begin{equation}}
\newcommand{\ee}{\end{equation}}
\newcommand{\ba}{\begin{aligned}}
\newcommand{\ea}{\end{aligned}}
\newcommand{\ben}{\begin{displaymath}}
\newcommand{\een}{\end{displaymath}}
\newcommand{\bea}{\begin{eqnarray}}
\newcommand{\eea}{\end{eqnarray}}
\newcommand{\bean}{\begin{eqnarray*}}
\newcommand{\eean}{\end{eqnarray*}}
\newcommand{\f}{\frac}
\newcommand{\p}{\partial}

\usepackage{amsmath,amssymb}
\usepackage{graphicx,color}

\renewcommand{\theequation}{\thesection.\arabic{equation}}

\def\l {\lambda}

\def\a {\alpha}

\def\b {\beta}

\def\d {\delta}
\def\e {\epsilon}
\def\s {\sigma}
\def\e {\epsilon}
\def\z {\zeta}

\def\m{\mu}
\def\n{\nu}

\definecolor{green}{rgb}{0,0.5,0}

\def\p{\partial}

\usepackage{epsfig}

\usepackage{xargs}
\usepackage[colorinlistoftodos,prependcaption,textsize=tiny]{todonotes}
\newcommandx{\Stefan}[1]{\todo[backgroundcolor=red!25,bordercolor=red,noline]{S:#1}}
\newcommandx{\George}[1]{\todo[backgroundcolor=blue!25,bordercolor=blue,noline]{G:#1}}

\numberwithin{equation}{section}

\begin{document}

\begin{titlepage}       \vspace{10pt}           \hfill {~~~~~~~~~~~~~~~}

\vspace{20mm}

\begin{center}

{\large \bf Canonical maps and integrability in $T\bar T$ deformed 2d CFTs\footnote{Prepared for 
`Integrability, Quantization, and Geometry - Dubrovin memorial volume',  edited by I. Krichever, S. Novikov, 
O. Ogievetsky and S. Shlosman}}

\vspace{30pt}

George Jorjadze$^{a,\,b}~$ and Stefan Theisen$^{c}$ 
\\[6mm]

{\small
{\it ${}^a$Free University of Tbilisi,\\
		Agmashenebeli Alley 240, 0159, Tbilisi, Georgia}\\[2mm]
{\it ${}^b$Razmadze Mathematical Institute of TSU,\\
Tamarashvili 6, 0177, Tbilisi, Georgia}\\[2mm]
{\it ${}^c$Max-Planck-Institut f\"ur Gravitationsphysik, Albert-Einstein-Institut,\\ 
14496, Golm, Germany}
}

\vspace{20pt}

\end{center}

\centerline{{\bf{Abstract}}}
\vspace*{5mm}
\noindent
We study $T\bar T$ deformations of 2d CFTs with periodic boundary conditions. We relate these systems to string models
on $\rr\times {S}^1\times{\cal M}$,
where $\cal M$ is the target space of a 2d CFT. 
The string model in the light cone gauge is identified with the corresponding 2d CFT and in the static gauge it 
reproduces its $T\bar T$  deformed system. This relates the deformed system and the initial one by a worldsheet
coordinate transformation, which becomes a time dependent canonical map in the Hamiltonian treatment. 
The deformed Hamiltonian defines the string energy and we express it in terms of the chiral Hamiltonians of the initial 2d CFT. 
This allows exact quantization of the 
deformed system, if the spectrum of the initial 2d CFT is known. The generalization to non-conformal 2d field 
theories is also discussed.

\vspace{15pt}

\end{titlepage}

\newpage

\section{Introduction}

The so-called $T\bar T$ deformation of two-dimensional quantum field theories, which was introduced  
by Zamolodchikov in 2004 \cite{Zamo}, has recently attracted much attention. Being a deformation by an irrelevant 
operator, one would naively expect that the deformed theory looses any of the nice properties the 
undeformed theory might have had and that the UV behaviour gets completely out of control. 
But this is not the case. For instance, in \cite{Zamo-Smirnov} it was shown that if the originial theory is integrable,
so is the deformed one. Another remarkable fact is that the spectrum of the deformed theory formulated 
on a cylinder can be determined exactly from the one of the undeformed theory \cite{Zamo,Zamo-Smirnov,Cavaglia}.    
   
An interesting observation first made in  \cite{Cavaglia}  is the connection between a deformed free boson and 
string theory.
More precisely, it was shown that the classical dynamics of the deformed system is that controlled by 
the Nambu-Goto action with three-dimensional Minkowski space as target space, after fixing the 
static gauge. In the same paper this was generalized to several free bosons and also to a single 
boson with an arbitrary potential. Further generalizations and refinements along these lines (and beyond) 
were considered in \cite{Bonelli}, \cite{Conti:2018jho}  and \cite{Freedman}, again at the classical level.  The relation between 
$T\bar T$ deformed CFTs and the quantum string was studied in detail in \cite{Verlinde}.
  
Here we also consider the connection between deformed field theories and 
string theory, mainly at the classical level. As a large part of our analysis will be within the Hamiltonian framework, 
the next section reviews  the Hamiltonian treatment of 
two-dimensional Lagrangian field theories. 
While the Lagrangian treatment is more 
familiar and transparent, the Hamiltonian one is more convenient for generalizations. 
The main examples are non-linear sigma-models with a metric and anti-symmetric tensor 
background. Classically they are always conformally invariant. Within the context 
of string theory one needs to impose conditions on the background fields, but this will not play a role in our 
classical discussion. A simple generalization, which explicitly breaks the conformal symmetry, is 
adding a potential. 

In Section 3 we look at the $T\bar T$ deformation of these theories, again in the Hamiltonian framework. 
A simple formula for the deformed Hamiltonian density for systems with symmetric 
canonical energy-momentum tensor can be derived. This formula is valid for arbitrary (classical) CFTs which are 
characterized by two independent components of the energy-momentum tensor whose Poisson brackets generate two 
copies of the centerless Virasoro algebra. 

The simplest conformally invariant sigma-model is a free massless scalar field on a cylinder. Its deformation will be 
reviewed in Section 4, with emphasis on the connection to closed string dynamics in three-dimensional space-time, 
where one spatial coordinate is compactified on $S^1$. When the latter is 
formulated in a diffeomorphism invariant way, the deformed free scalar is obtained by breaking the 
invariance through fixing the static gauge. This gauge identifies the time and one spatial coordinate of the 
target space with the worldsheet coordinates.
For this reason compactification is necessary. 
The string energy is, up to an additive constant, equal to the 
Hamiltonian of the deformed theory.  
If one  chooses light-cone gauge instead, one reaches the undeformed 
theory. We generalize the light-cone gauge treatment of a closed string dynamics with a compactified 
spatial coordinate, using as space-time light-cone directions those of the cylinder. 
This generalization is straightforward. In particular, in this gauge the string energy can be computed explicitly 
and by using its gauge 
invariance one obtains an expression for the Hamiltonian -- rather than the density --  of the deformed theory 
in terms of the Hamiltonian  of the undeformed theory. 
This result applies, in fact, 
to more general undeformed theories than just the free massless scalar.

The relation between the deformed and the undeformed theory as simply choosing different gauges in the 
string theory,  implies that the undeformed and the deformed theory are related by a (time-dependent) 
canonical transformation. This will be shown in detail. 
The worldsheet coordinate transformation between the two gauges depends on the solutions 
of the equation of motion in the fixed gauge. We use the explicit form of this transformation 
to obtain the Hamiltonian of the deformed theory without resorting 
to the gauge invariance of the string energy.   
 
In Section 5 we show how the previous discussion extends to general conformally invariant 
sigma models and to the case when one adds a potential.  
A remarkable example here is the Liouville model with a negative cosmological constant. 
We show that the corresponding string model is the SL(2,$\rr)$ WZW theory with vanishing stress 
tensor \cite{Sundborg:2013bya}. This string model in the static and light-cone gauges coincides to the $T\bar T$ deformed 
and the initial Liouville models, respectively. 
  
Some of the results reported in this note were obtained but not published
about two years ago \cite{J-T} and they have meanwhile appeared in various papers.  
We have taken the opportunity of being asked to contribute to this volume to 
include them, with due reference to the existing literature. 
Most importantly we point out 
\cite{Verlinde, Sfondrini1,Frolov1,Frolov2,Sfondrini2, Tateo} for extensive discussions of the relation between the $T\bar T$ 
deformed and the initial 2d field theories in the context of worldsheet gauge transformations.


\section{Hamiltonian formulation of 2d field theory}

We consider two-dimensional classical field theories on a cylinder with circumference $2\pi$, described by an action
\be\label{Action_L}
S[\phi]=\f{1}{2\pi}\int\mbox{d}\tau\,\mbox{d}\s \,\,{\cal L}(\phi, \dot\phi, \acute\phi)~.
\ee
Here, $\tau$ and $\s$ are time and space coordinates, respectively, $\phi:=(\phi^1,\dots , \phi^N)$ denotes a set of 
periodic fields, $\phi(\tau,\s+2\pi)=\phi(\tau,\s)$, and 
we use the notation  $\dot \phi:=\p_\tau \phi$, $\acute \phi:=\p_\s \phi$. 

The components of the canonical stress tensor ($a,b\in\{\tau,\s\}$)
\be\ba\label{Canonical T}
&T^{\,a}_{\,~b}={\p{\cal L}\over\p(\p_a\phi^k)}\p_b\phi^k-\delta^a_{~b}\,{\cal L}
\ea\ee
satisfy, by Noether's theorem, the local conservation laws
\be\label{dT=0}
\p_a T^{\,a}_{\,~b}=0\,.
\ee
The first order formulation of the same dynamics is obtained from the action
\be\label{Action_H}
S[\Pi, \phi]=\int\mbox{d}\tau\int_0^{2\pi} \f{\mbox{d}\s}{2\pi} \left[\Pi_k\,\dot \phi^k-{\cal H}(\Pi, \phi,\acute \phi)\right]~,
\ee
where $\Pi_k$ are the periodic canonical momenta, 
$\Pi(\tau,\s+2\pi)=\Pi(\tau,\s)$.
We assume that the Lagrangian in \eqref{Action_L} is non-singular,\footnote{Singular Lagrangians also lead to 
the action \eqref{Action_H} by Hamiltonian reduction, 
but with a reduced number of target space fields.} i.e.  the velocities 
$\dot \phi^k$ are solvable in terms of the momenta $\Pi_k$. 

The stress tensor components \eqref{Canonical T} are
\be\ba\label{Canonical T_H}
&T^{\,\tau}_{~\,\tau}={\cal H}~,  &&T^{\,\tau}_{\,~\s}=\Pi_{k}\,\acute \phi^{\,k}~,
\\[1mm]
&T^{\,\s}_{\,~\tau} =-\f{\p{\cal H}}{\p \Pi_{k}}\,\f{\p{\cal H}}{\p\acute \phi^{\,k}}~,  \qquad    
&&T^{\,\s}_{\,~\s} ={\cal H}-\Pi_k\,\f{\p{\cal H}}{\p \Pi_{k}}-\acute \phi^{\,k}\,\f{\p{\cal H}}{\p\acute \phi^{\,k}}~,
\ea\ee
and the conservation laws \eqref{dT=0} follow from the Hamilton equations of motion
\be\label{Hamilton eq}
\dot \phi^k= \f{\p\cal H}{\p \Pi_k}~, \qquad  \dot \Pi_k= -\f{\p\cal H}{\p \phi^k}+\p_\s\left(\f{\p\cal H}{\p\acute \phi^{\,k}}\right)~.
\ee
Note that the covariant canonical stress tensor $T_{ab}$ in 2d Minkowski space is symmetric ($T_{\tau\,\s}=T_{\s\,\tau}$) 
when the Hamiltonian density satisfies the  condition
\be\label{symmetry condition}
\f{\p{\cal H}}{\p \Pi_{k}}\,\f{\p{\cal H}}{\p\acute \phi^{\,k}}=\Pi_k\,\acute \phi^{\,k}\,.
\ee
Below we assume that \eqref{symmetry condition} is fulfilled, without referring to 2d metric structure.\footnote{While we 
can always add improvement terms to symmetrize the energy-momentum tensor, here we 
assume that the canonical one is symmetric.} 

We also assume that the canonical stress tensor \eqref{Canonical T_H} is traceless, i.e.  
\be\label{traceless condition}
\hat{V}[{\cal H}]=2\,{\cal H}~, \qquad \text{where} ~~~
\hat V=\Pi_k\,\f{\p}{\p \Pi_k}+\acute \phi^{\,k}\,\f{\p}{\p\acute \phi^{\,k}}~.
\ee
In this case
\be\label{T=H,P}
T^{\,a}_{\,~b}=\left(\begin{array}{cr}{\cal H} &{\cal P}\\-{\cal P} &-{\cal H}
\end{array}\right)~,
\qquad \text{with} \quad {\cal P}:=\Pi_{k}\,\acute \phi^{\,k}~.
\ee

The components $T^{\,\tau}_{~\tau}={\cal H}$ and $T^{\,\tau}_{~\s}={\cal P}$ are interpreted as the energy and the 
momentum densities, respectively. They obey the Poisson bracket relations
\be\ba\label{P,H-PB}
&\{{\cal P}(\s_1), {\cal P}(\s_2)\}=\{{\cal H}(\s_1), {\cal H}(\s_2)\}=2\pi\big[{\cal P}(\s_1)+{\cal P}(\s_2)\big]\d'(\s_2-\s_1), \\[1mm]
&\{{\cal P}(\s_1), {\cal H}(\s_2)\}=\{{\cal H}(\s_1), {\cal P}(\s_2)\}=2\pi\big[{\cal H}(\s_1)+{\cal H}(\s_2)\big]\d'(\s_2-\s_1),
\ea\ee
which follow from  the canonical Poisson brackets, 
\be\label{Canonical PB 0}
\{\Pi_k(\s_1), \phi^{l}(\s_2)\}=2\pi\,\d_{k}^{\,\,l}\,\d(\s_1-\s_2)~,
\ee 
and the conditions \eqref{symmetry condition} and \eqref{traceless condition}.
The Lie algebra \eqref{P,H-PB} is equivalent to 
\be\ba
&\{T(x), T(y)\}=2\pi\big[T(x)+T(y)\big]\d'(y-x)\,,& \quad &\{T(x), \bar T(\bar x)\}=0\,,\\[1mm] \label{PB T-bar T}
&\{\bar T(\bar x),\bar T(\bar y)\}=
2\pi\big[\bar T(\bar x)+\bar T(\bar y)\big]\d'(\bar y-\bar x)\,,&  &
\ea\ee
with
\be\label{T-bar T}
T(x)=\f{1}{2}\big[{\cal H}(x)+{\cal P}(x)\big]~, \qquad 
\bar T(\bar x)=\f{1}{2}\big[{\cal H}(-\bar x)-{\cal P}(-\bar x)]~.\ee 
The conservation laws \eqref{dT=0} in terms of $T$ and $\bar T$ become
\be\label{chiral T}
\p_{\bar x}T=0~, \qquad \p_{x}\bar T=0~,
\ee
where $x=\tau+\s$ and $\bar x=\tau-\s$ are the chiral coordinates, and  
we arrive at the standard formulation of 2d CFT with zero central charge.

In a more general treatment, a 2d CFT on a cylinder is provided by two periodic functions
$T(x)$ and $\bar T(\bar x)$, which satisfy the Poisson bracket relations \eqref{PB T-bar T},
without referring to the canonical structure \eqref{Action_H}. 
Thus, the Hamiltonian density $\cal H$ that satisfies the conditions \eqref{symmetry condition} and \eqref{traceless condition} 
corresponds to a classical 2d CFT.

A standard example is the $\s$-model
\be\label{Action_sigma}
S_{_{G,B}}[\phi]=\f{1}{4\pi}\int\mbox{d}\tau\,\mbox{d}\s \left[\dot \phi^k\, G_{kl}(\phi)\, \dot \phi^l
-\acute \phi^{\,k}\,G_{kl}(\phi)\,\acute \phi^{\,l}-2\dot \phi^k\, B_{kl}(\phi)\,\acute \phi^{\,l}\right]~,
\ee
where $G_{kl}(\phi)$ is a target space metric tensor and $B_{kl}(\phi)$ is a 2-form on the target space. 
This system has stress tensor 
\be\ba\label{T^m_n}
&T^{\,\tau}_{~\,\tau}=-T^{\,\s}_{~\,\s}=\f{1}{2}\left(\dot \phi^k\, G_{kl}\, \dot \phi^l+ \acute \phi^{\,k}\,G_{kl}\,\acute \phi^{\,l}\right), 
\quad  &&T^{\,\tau}_{~\,\s}=-T^{\,\s}_{~\,\tau}=\dot \phi^{\,k}\, G_{kl}\,\acute \phi^{\,l} ~,
\ea\ee
and Hamiltonian density 
\be\label{Sigma model H}
{\cal H}_{_{G,B}}=\f{1}{2}\left[\Pi_k\, G^{kl}\, \Pi_l+\acute \phi^{\,k}\left(G_{kl}-B_{km}\,G^{mn}\,B_{nl}\right)\acute \phi^{\,l}\right]
+\Pi_k\,G^{kj}\,B_{jl}\,\acute \phi^{\,l}~, 
\ee
which indeed satisfies conditions \eqref{symmetry condition} and \eqref{traceless condition}.

Adding a potential $U(\phi)$ to a 2d CFT
\be\label{H with U(q)}
\tilde{\cal H}={\cal H}+U(\phi)~,
\ee
leads to a stress tensor with non-zero trace
\be\label{T=H0,P,U}
T^{\,a}_{\,~b}=\left(\begin{array}{cr}{\cal H}+U(\phi) &{\cal P}~~~~~~~\\-{\cal P} &-{\cal H}+U(\phi)
\end{array}\right)~.
\ee

\section{$T\bar T$ deformation of 2d Hamiltonian systems}

The following analysis is usually done in the Lagrangian formulation (cf. e.g. \cite{Cavaglia,Bonelli,Tateo}). 
Here we present a Hamiltonian version of these well-known results.  

We introduce the $T\bar T$ deformation of the system \eqref{Action_H} as \cite{Zamo}
\be\label{Def-Action_H} 
S_\a[\Pi,\phi]=\int\mbox{d}\tau\int_{0}^{2\pi}\f{\mbox{d}\s}{2\pi} \left[\Pi_k\,\dot \phi^k-{\cal H}_\a(\Pi, \phi,\acute \phi)\right]~,
\ee
with ${\cal H}_\a$ defined by the `initial' condition ${\cal H}_0={\cal H}$ and the differential equation
\be\label{Def-eq}
\f{\p{\cal H}_\a}{\p\a}=\f 1 2 \mbox{det}[T_{(\a)}]\,.
\ee
Here $T_{(\a)}^{\,\,a}\,_b $ is the canonical stress tensor obtained from \eqref{Canonical T_H} 
by the replacement ${\cal H}\mapsto {\cal H}_\a$. 

Note that $\mbox{det}[T^{a}_{~\,b}]={\cal P}^2-{\cal H}^2=-4 T\bar T$ for a 2d CFT. 
Thus, the first order correction to the Hamiltonian density of a 2d CFT is
\be\label{First orde H_a}
{\cal H}_\a={\cal H}-2\,\a\, T\bar T +\cdots~;
\ee
hence the name $T\bar T$ deformation. However, the higher order terms do not have this structure and 
are more complicated. 

From \eqref{Def-eq} and \eqref{Canonical T_H} follows that  ${\cal H}_\a$ satisfies the equation 
\be\label{general equation for H_a}
2\f{\p{\cal H}_\a}{\p\a}={\cal H}_\a^2-{\cal H}_\a\,\hat{V}\left[{\cal H}_\a\right]+
{\cal P}\,\f{\p{\cal H}_\a}{\p \Pi_{\,k}}\,\f{\p{\cal H}_\a}{\p\acute \phi^{\,k}}~,
\ee
and one is looking for solutions which are analytic in $\a$ at $\a=0$.

Using \eqref{general equation for H_a}, one shows by a straightforward but slightly tedious calculation that the variable 
$Y_\a=\f{\p{\cal H}_\a}{\p \Pi_{\,k}}\,\f{\p{\cal H}_\a}{\p \acute \phi^{\,k}}-{\cal P}$ satisfies the equation
\be\label{Eq for Y_a}
\f{\p Y_\a}{\p\a}={\cal H}_\a\,Y_\a-\f{1}{2}\,\hat{V}\left({\cal H}_\a\,Y_\a\right)+
\f{1}{2}\,{\cal P}\left(\f{\p{\cal H}_\a}{\p \Pi_k}\,\f{\p Y_\a}{\p \acute \phi^k}
+\f{\p{\cal H}_\a}{\p \acute \phi^{\,k}}\,\f{\p Y_\a}{\p \Pi_{\,k}}\right)~.
\ee
From the `initial' condition $Y_{\a=0}=0$ then follows that $Y_\a$ remains zero for all $\a$.
Hence, ${\cal H}_\a$ satisfies the condition
\be\label{symmetry condition-a}
\f{\p{\cal H}_\a}{\p \Pi_{\,k}}\,\f{\p{\cal H}_\a}{\p \acute \phi^{\,k}}=\Pi_k\,\acute \phi^{\,k}\,,
\ee
and \eqref{general equation for H_a} reduces to
\be\label{reduced equation for H_a}
2\,{\p_\a{\cal H}_\a}={\cal H}_\a^2-{\cal H}_\a\,\hat{V}\left[{\cal H}_\a\right]+
{\cal P}^2~.
\ee
This equation can be easily integrated if the stress tensor of the undeformed theory is traceless. 
Indeed, taking into account \eqref{traceless condition} and  $\hat V[{\cal P}]=2\cal P$, one finds that ${\cal H}_\a$ is 
expressed in terms of $\cal H$ and $\cal P$ only. Dimensional analysis suggests the ansatz  
\be\label{Ansatz} 
{\cal H}_\a=F_\a(r\,{\cal H}+\a\,{\cal P}^2) ~,
\ee
where $r$ is a real number. Inserting it into 
\eqref{symmetry condition-a} one finds $F'(u)=\left({r^2+4\,\a\, u}\right)^{-\f{1}{2}}$.
Integration, requiring the regularity condition at $\a=0$ and  that it satisfies \eqref{reduced equation for H_a}, leads to \cite{Tateo}
\be\label{H_a}
{\cal H}_\a=\f{1}{\a}\,\left(\sqrt{1+2\,\a\, {\cal H}+\a^2\,{\cal P}^{2}}-1\right)~.
\ee

The structure of the energy-momentum tensor of the deformed theory is 
\be\label{Deformed T}
T_{(\a)\,b}^{\,\,a}=\left(\begin{array}{cr}{\cal H}_\a &{\cal P}\\-{\cal P} &-{\cal K}_\a
\end{array}\right),
\ee
with
\be\label{T-T relation}
{\cal K}_\a={1\over\a}\left({1-\a^2\,{\cal P}^2\over \sqrt{1+2\,\a\,{\cal H}+\a^2{\cal P}^2}}-1\right)
=\f{{\cal H}_\a+\a\,{\cal P}^2}{1+\a\,{\cal H}_\a}\,.
\ee
One also verifies 
\be\label{Tr=a Det} 
 \mbox{Tr}[T_{(\a)}]=-\a\,\mbox{det}[T_{(\a)} ]~
\ee
and, therefore, for a 2d CFT, ${\cal H}_\a$ satisfies the linear equation 
\be\label{Eq with Tr}
2\,\a\,\p_\a{\cal H}_\a+2\,{\cal H}_\a-\hat{V}[{\cal H}_\a]=0~.
\ee

The above results, in particular the form of the deformed Hamiltonian density \eqref{H_a}, 
were derived for a particular class of conformal field theories, but one wonders how general they are. 
If we assume that the energy-momentum tensor of the undeformed theory is 
symmetric, it has only two independent components, $T$ and $\bar T$. In terms of those 
\be\label{H_a2}
{\cal H}_\a=\f{1}{\a}\left(\sqrt{1+2\,\a\,(T+\bar T)+\a^2(T-\bar T)^2}-1\right)\,.
\ee
Using the algebra \eqref{PB T-bar T}, which holds for any CFT, one verifies that 
\be\label{Hamiltonian H_a}
\dot{\cal H}_\a=\{H_\a,{\cal H}_\a\}=\p_\s(T-\bar T),\qquad\hbox{where}\qquad
H_\a=\int_0^{2\pi} \f{{\text{d}\s}}{2\pi}\,{\cal H}_\a\,.
\ee
Imposing the $\tau$-component of the conservation equation in \eqref{dT=0} for 
the deformed theory, this shows that $T^{\,\s}_{\,~\tau}=\bar T-T$ is not deformed. Imposing 
instead the $\s$-component and requiring symmetry of $T_{(\a)}$ leads to 
\be
T_{(\a)\,\s}^{\,\s}={\cal H}_\a-2\,\f{\p{\cal H}_\a}{\p T}T-2\f{\p{\cal H}_\a}{\p\bar T}\bar T~.
\ee
These results are completely general for two-dimensional conformal field theories, in particular the expression 
\eqref{H_a2} for the Hamiltonian density. 

We stress that our discussion so far was classical. In particular, in the quantized theory the algebra \eqref{PB T-bar T} is modified 
by a central extension leading to the Virasoro algebra. 
Even for string theory, when the contribution of the 
ghosts is included, the above calculation does not go through straightforwardly because of ordering 
issues in the expression for ${\cal H}_\a$.

The $T\bar T$ deformation of the model \eqref{H with U(q)}, 
with the potential $U(\phi)$, can be performed similarly.
In this case  
$\hat V[\tilde{\cal H}]=2{\cal H}$ and $\tilde{\cal H}_\a$ 
becomes a function of
${\cal H}$, $\cal P$ and $U(\phi)$ only. Repeating the arguments which lead to 
\eqref{H_a}, we obtain \cite{Bonelli}
\be\label{H_a with U(q)}
\tilde{\cal H}_\a=\f{1}{\b}\,\left[\sqrt{1+2\,\b \,{\cal H}+\b^2\,{\cal P}^{2}}+\f{\a\, U(\phi)}{2}\right]-\f{1}{\a}~,
\ee
with 
\be\label{beta=}
\b=\a\left(1-\f{\a}{2}\,U(\phi)\right)~.
\ee
The check of \eqref{symmetry condition-a} and \eqref{Def-eq}  is again straightforward.

\section{Integrability of the deformed 2d massless free field}

In this section we investigate integrability of the deformed  massless free-field model with the undeformed Lagrangian
\be\label{FF-lagrangian}
{\cal L}=\f{1}{2}\left(\dot \phi^2-\acute \phi^{\,2}\right).
\ee
The energy and momentum densities 
\be\label{FF-H}
{\cal H}=\f{1}{2}\left(\Pi^2+\acute \phi^{\,2}\right), \qquad {\cal P}=\Pi\,\acute \phi~, 
\ee
lead to the following deformed Hamiltonian density 
\be\label{D-FF H} 
{\cal H}_\a=\f{1}{\a}\left(\sqrt{1+\a\left(\Pi^2+\acute \phi^{\,2}\right)+\a^2 \Pi^2\,\acute \phi^{\,2}}-1\right).
\ee
From the related Lagrangian
\be\label{D-FF L} 
{\cal L}_\a=-\f{1}{\a}\left(\sqrt{1+\a\,\acute \phi^{\,2}-\a\,\dot \phi^2}-1\right)~,
\ee
one derives a non-linear dynamical equation 
which is hard to integrate directly. 
Furthermore the construction of the Hamilton operator by \eqref{D-FF H} seems a 
highly nontrivial problem due to the non-polynomial dependence of ${\cal H}_\a$ on the canonical variables.
However, the deformed free-field theory is related to a 3d string with one compactified coordinate \cite{Cavaglia}.
This enables us to integrate the system both at classical and quantum levels. 
We first consider the Lagrangian approach to the compactified 3d string dynamics and then turn to its Hamiltonian treatment. 

For later use we note that $\Pi$ and $\dot \phi$ of the deformed theory \eqref{D-FF L} are related by
\be\label{p and dot q}
\Pi=\f{\dot \phi}{\sqrt{1+\a\,\acute \phi^{\,2}-\a\,\dot \phi^2}}~, \qquad \dot \phi=\Pi\sqrt{\f{1+\a\,\acute \phi^{\,2}}{1+\a\,\Pi^2}}~,
\ee
and the energy and momentum densities in the Lagrangian formulation become
\be\label{cal H_a and P}
{\cal H}_\a=\f{1}{\a}\left(\f{1+\a\,\acute \phi^{\,2}}{\sqrt{1+\a\,\acute \phi^{\,2}-\a\,\dot \phi^2}}-1\right), 
\qquad {\cal P}=\f{\dot \phi\,\acute \phi}{\sqrt{1+\a\,\acute \phi^{\,2}-\a\,\dot \phi^2}}~.
\ee

\subsection{Lagrangian approach to a compactified 3d string}

We start with a review of the connection between the string and the deformed system \cite{Cavaglia}.  
The Nambu-Goto action for a closed string is
\be\label{Nambu-Goto action}
S=-\f{1}{2\pi\a}\int \mbox{d}\tau \int_0^{2\pi}\mbox{d}\s \sqrt{(\dot X\,\acute X)^2-(\dot X\,\dot X)(\acute X\,\acute X)}~.
\ee
$X:=(X^0, X^1, X^2)$ is a vector in 3d Minkowski space and $1/\a$ is proportional to the string tension.
We use the notation $(X X)=X^\mu X^\nu g_{\mu\nu}$ 
with the target space metric tensor $g_{\m\n}=\mbox{diag}(-1,1,1)$. 
This theory has two-dimensional diffeomorphism invariance and is classically equivalent to the 
Polyakov action with a world-sheet metric.

To connect the deformed free-field theory to the closed string dynamics, we
compactify the coordinate $X^1$  on the unit circle and consider string configurations with winding number one around 
this circle, i.e. we identify $X^1\simeq X^1+2\,\pi$. 
This enables us to parameterize $X^1$ by $\s$. 
We then identify $X^0$ with $\tau$ and  parameterize $X^2$ by $\sqrt{\a}\,\phi$, i.e.
we use the static gauge where
\be\label{static gauge}
X^\m=\left(\begin{array}{c}\tau\\ \s \\ \sqrt{\a}\,\phi  \end{array}\right)~,\qquad
\dot X^\m=\left(\begin{array}{c}1\\0\\ \sqrt{\a}\,\dot \phi  \end{array}\right)~,\qquad
\acute X^{\,\m}=\left(\begin{array}{c}0\\ 1 \\ \sqrt{\a}\,\acute \phi\end{array}\right)~.
\ee
In this gauge the string Lagrangian in \eqref{Nambu-Goto action} reduces to the deformed Lagrangian \eqref{D-FF L}, 
up to the additive constant $1/\a$.

The string energy-momentum densities obtained from the Nambu-Goto action \eqref{Nambu-Goto action},
\be\label{P_m}
{\cal P}^\m=
\f{1}{\a}\,\f{\dot X^\m(\acute X\,\acute X)-\acute X^{\m}(\dot X\,\acute X)}{\sqrt{(\dot X\,\acute X)^2
-(\dot X\,\dot X)(\acute X\,\acute X)}}~,
\ee
satisfy the (primary) constraints 
\be
(\acute X\,{\cal P})=0\,,\qquad \a^2\,(\acute X\,\acute X)+({\cal P}\,{\cal P})=0\,.
\ee
As in the uncompactified case, the tangent vectors $\acute{X}$  and  $\dot{X}$ are assumed  
spacelike and  timelike, respectively, and $X^0$ is monotonically increasing  in $\tau,$
i.e. 
\be\label{tangent vectors} 
(\acute X\,\acute X)>0~, \qquad (\dot X\,\dot X)<0~, \qquad \dot X^0>0~.
\ee
The momentum density  ${\cal P}^\m$ is then timelike and ${\cal P}^0$ is positive.
In static gauge
\be\ba\label{P_m in static gauge}
&{\cal P}^0=\f{1}{\a}\,\f{1+\a\,\acute \phi^{\,2}}{\sqrt{1+\a\,\acute \phi^{\,2}-\a\,\dot \phi^2}}~,\\
& {\cal P}^1=\f{-\dot \phi\,\acute \phi}{\sqrt{1+\a\,\acute \phi^{\,2}-\a\,\dot \phi^2}}~,\qquad
{\cal P}^2=\f{1}{\sqrt{\a}}\,\f{\dot{\phi}}{\sqrt{1+\a\,\acute \phi^{\,2}-\a\,\dot \phi^2}}~.
\ea\ee
Comparing these expressions to \eqref{p and dot q}-\eqref{cal H_a and P}, we find 
\be\label{P_m=cal P}
{\cal P}^0={\cal H}_\a+\f{1}{\a}~, \qquad {\cal P}^1=-{\cal P}~,\qquad {\cal P}^2=\f{1}{\sqrt{\a}}\, \Pi~.
\ee
Integrating the densities over $\s$ gives the gauge invariant 
string energy-momentum. In particular, the string energy reads
\be\label{string energy}
E_{\text{str}}= \int_0^{2\pi}\f{\text{d}\s}{2\pi}\,{\cal P}^0(\s)=H_\a +\f{1}{\a}~,
\ee
where $H_\a$ is the energy of the deformed system \eqref{Hamiltonian H_a}.

Thus, the deformed system \eqref{D-FF L} and the compactified 3d string in the gauge 
\eqref{static gauge} are identical dynamical systems.
On the other hand, it is well known that the classical string dynamics is integrable in the light-cone gauge. 
The compactification of 
the coordinate $X^1$ does not destroy integrability, but rather modifies it, as we show below.

The static gauge \eqref{static gauge} is not a conformal one for which 
one requires $(\dot X\,\acute X)=0$ 
and $(\dot X\,\dot X)+(\acute X\,\acute X)=0$ and the equation of motion for $X^\m$ becomes the free 
wave equation.  These two constraints have to be imposed on the solutions.  
We denote the conformal worldsheet coordinates by $(\tau_c,\s_c)$, to distinguish them from $(\tau, \s)$, and 
introduce the corresponding chiral coordinates $z=\tau_c+\s_c$ and $\bar z=\tau_c-\s_c$.
One then has $\p_z\p_{\bar z} X^\m=0$, and its solutions 
\be\label{CG solution}
X^\m=\Phi^\m(z)+\bar\Phi^\m(\bar z)
\ee
are restricted to satisfy the conformal gauge conditions
\be\label{CG conditions}
(\Phi'\, \Phi')=0~, \qquad (\bar \Phi'\,\bar \Phi')=0~.
\ee
The chiral functions $\Phi'^{\,\m}(z)$ and $\bar \Phi'^{\,\m}(\bar z)$ are periodic. 
Therefore, similarly  to the uncompactified case, $\Phi^\m(z)$ and $\bar \Phi^\m(\bar z)$  
obey the monodromy conditions
\be\label{mode of Phi^mu}
\Phi^\m(z+2\pi)=\Phi^\m(z)+2\pi\,\rho^\m~, \qquad 
\bar \Phi^\m(\bar z+2\pi)=\bar \Phi^\m(\bar z)+2\pi\,\bar \rho^\m~,
\ee
where $\rho^\m$ and $\bar \rho^\m$ are the zero modes of  $\Phi'^{\,\m}(z)$ and $\bar \Phi'^{\,\m}(\bar z)$, respectively.
From the periodicity conditions in $\s$ one finds
\be\label{P=bar P}
\rho^0=\bar\rho^{\,0}~, \qquad \rho^1=\bar\rho^{\,1}+L~, \qquad \rho^2=\bar\rho^{\,2}~,
\ee 
where $L$ is the winding number around the compactified coordinate $X^1$. For now we analyze the case of 
general $L$, though our interest is $L=1$.

To find independent variables on the constraint surface \eqref{CG conditions}, we follow the standard scheme
and introduce the light-cone coordinates  $X^\pm=X^0\pm X^1$. Note that while one usually chooses the 
space-time light-cone directions along two non-compact coordinates, our definition of $X^\pm$ involves the 
compact direction $X^1$.  
The remaining freedom of conformal transformations 
allows us to simplify the chiral components of $X^+$ as in the uncompactified 
case\footnote{The conditions \eqref{Phi^+ map} require $\rho^+>0$ and $\bar{\rho}^+>0$. We will see in \eqref{rho^+=} that 
these conditions are indeed fulfilled.}
\be\label{Phi^+ map}
\Phi^+(z)=\rho^+ z~,  \qquad \bar \Phi^+=\bar \rho^{\,+} \bar z~.
\ee
The constraints \eqref{CG conditions} can then be written as 
\be\label{F^-=}
\rho^+ \,\Phi'{\,^-}(z)={\a}\, F'^{\,2}(z)~, \qquad \bar \rho^{\,+}\,\bar\Phi'{\,^-}(\bar z)={\a}\,\bar F'^{\,2}(\bar z)~,
\ee
where $X^2$ is rescaled similarly to \eqref{static gauge}, i.e. $\Phi^2(z)=\sqrt{\a}\,F(z)$ and 
$\bar\Phi^2(\bar z)=\sqrt{\a}\,\bar F(\bar z)$.

As a result, one obtains the following parameterization of the string coordinates
\be\label{l-c gauge solution}
X^\m=\left(\begin{array}{c} \f{1}{2}\left[\rho^+z +\Phi^-(z)+\bar \rho^{\,+}\bar z+
\bar\Phi^{\,-}(\bar z)\right]\\[2mm]
\f{1}{2}\left[\rho^+z - \Phi^-(z)+\bar \rho^{\,+}\bar z-\bar\Phi^{\,-}(\bar z) \right]\\[2mm]
\sqrt{\a}\left[ F(z)+\bar F(\bar z)\right]
\end{array}\right).
\ee
The functions $F(z)$ and $\bar F(\bar z)$ have the mode expansions 
\be\label{modes of F}
F(z)=\f{q+pz}{2}+\f{\mathrm{i}}{\sqrt{2}}\sum_{m\neq 0}\f{a_n}{n}\,\text{e}^{-\mathrm{i}nz}, \qquad 
\bar F(\bar z)=\f{q+p\bar z}{2}+\f{\mathrm{i}}{\sqrt{2}}\sum_{n\neq 0}\f{\bar a_n}{n}\,\text{e}^{-\mathrm{i}n\bar z},
\ee
with $p=\f{2}{\sqrt{\a}}\,\rho^2$, and $\Phi^-(z)$ and $\bar\Phi^{\,-}(\bar z)$  are obtained from \eqref{F^-=} (see Appendix A). 
In particular, one has
\be\label{rho^-=}
\rho^-=\a\,\f{h}{\rho^+}~, \qquad \bar \rho^{\,-}=\a\,\f{\bar h}{\bar \rho^{\,+}}~, 
\ee
where $h$ and $\bar h$ are the chiral free-field Hamiltonians 
\be\label{H and bar H}
h=\int_0^{2\pi} \f{\mbox{d}z}{2\pi}\, F'^{\,2}(z)=\f{p^2}{4}+\sum_{n >0} |a_n|^2\,,~~~
\bar h=\int_0^{2\pi} \f{\mbox{d}\bar z}{2\pi}\, \bar F'^{\,2}(\bar z)=\f{p^2}{4}+\sum_{n> 0} |\bar a_n|^2\,.
\ee
Note that we set $\bar p=p$ in \eqref{modes of F}, due to the third
relation in \eqref{P=bar P}. The other two relations of \eqref{P=bar P}, in terms of the light-cone variables, read
\be\label{monodromy}
\rho^+ + \rho^- - \bar \rho^{\,+}- \bar \rho^{\,-}=0~,  \qquad \rho^+ - \rho^- - \bar \rho^{\,+}+ \bar \rho^{\,-}=2L~.
\ee
For $L\neq 0$ this leads to differences for the compactified case as compared to the non-compact one.  

Indeed,  for $L=0$, the solution of \eqref{rho^-=}-\eqref{monodromy} is 
\be\label{n=0 solution}
 \rho^+=\bar \rho^{\,+}~,\qquad \rho^-=\bar \rho^{\,-}=\a\,\f{h}{\rho^+}=\a\,\f{\bar h}{\bar \rho^{\,+}}~, \qquad h=\bar h~.
\ee
Here, $\rho^+$ is a free dynamical variable. The condition $h=\bar h$ 
becomes, after quantization, the level matching condition in the zero winding sector. 

When $L\neq 0$,  we obtain instead the following solution of \eqref{rho^-=}-\eqref{monodromy} 
\be\ba\label{rho^+=}
&\rho^\pm=\f{1}{2}\left(\a\,{\cal E}_{L}\pm\f{\a}{L}(\bar h-h)\pm L\right),\quad
&\bar \rho^{\,\pm}=\f{1}{2}\left(\a\,{\cal E}_{L}\pm\f{\a}{L}(\bar h-h)\mp L\right),
\ea\ee
\be\label{string energy n} 
\text{with} ~~~~~~~~{\cal E}_{L}=\f{1}{L\,\a\,}\sqrt{L^4+2L^2\,\a(h+\bar h)+\a^2(h-\bar h)^2}~.
\ee
Here, solving quadratic equations, we choose the positive roots, since they correspond to the physical solutions 
for which $\rho^\pm>0$ and $\bar{\rho}^{\,\pm}>0$. 

Thus, for $L\neq 0$, the string solutions \eqref{l-c gauge solution} are completely 
parametrized by the chiral free fields $F(z), \,\bar F(\bar z)$. We now find
that the level matching condition is modified to 
\be
L(\rho^1+\bar\rho^1)=\a\,(\bar h-h)\,.
\ee

According to \eqref{P_m},  the string energy density in the conformal gauge is given by 
$\f{1}{\a}\,\p_{\tau_c} X^0$,
and from \eqref{rho^+=} we obtain 
the string energy for winding number $L$ 
\be\label{string energy n=} 
E_{\text{str}}^{(L)}=\f{1}{2\,\a}\left(\rho^+ + \rho^- +\bar \rho^{\,+}+\bar \rho^{\,-}\right)={\cal E}_L ~.
\ee
For winding number one, which corresponds to the deformed system, this yields
\be\label{string energy n=1} 
E_{\text{str}}=\f{1}{\a}\sqrt{1+2\,\a(h+\bar h)+\a^2(h-\bar h)^2}~,
\ee
and, due to the gauge invariance of the string energy, we obtain from \eqref{string energy}
\cite{Conti:2018jho} 
\be\label{deformed energy}
H_\a=\f{1}{\a}\left(\sqrt{1+2\,\a(h+\bar h)+\a^2(h-\bar h)^2}-1\right)~.
\ee 

This expression for the Hamiltonian should be contrasted with \eqref{H_a2}. There the Hamiltonian density of the deformed 
theory was expressed in terms of the energy-momentum densities of the undeformed theory while here 
the relation is between the integrated densities. Furthermore, this expression can be easily quantized as
$h$ and $\bar h$ are diagonal in the Fock-space of the undeformed theory. 

In Section 5.1 we will briefly discuss generalizations to general CFTs. In this case the expression for $H_\a$ is 
straightforwardly generalized by replacing $(h,\bar h)$ by $(L_0,\bar L_0)$ of the undeformed theory. In fact, many of the 
expressions in the following discussion are generalized if one replaces in the expression in Appendix A 
the $L_n$ of the free field by the generators of the Virasoro algebra of a general CFT. 

In Appendix B we derive \eqref{deformed energy} directly (without referring to the gauge invariance),
using the map that relates the worldsheet coordinates and the fields in two different gauges. We will now analyze 
this map in detail.

Comparing the string coordinates in the gauges \eqref{static gauge} and \eqref{l-c gauge solution},
we find the map from the coordinates $(z, \bar z)$ to $(\tau, \s)$\footnote{Recall that 
$z=\tau_c+\s_c$ and $\bar z=\tau_c-\s_c$.} 
\be\ba\label{s,t map}
\tau= \f{1}{2}\left[\rho^+z +\Phi^-(z)+\bar \rho^{\,+}\bar z+\bar\Phi^{\,-}(\bar z)\right]~,\\[1mm] 
\s= \f{1}{2}\left[\rho^+z -\Phi^-(z)+\bar \rho^{\,+}\bar z-\bar\Phi^{\,-}(\bar z)\right]~,
\ea\ee
and we also express the solutions of the deformed system by the undeformed one
\be\label{q=F+bar F}
\phi(\tau,\s)= F(z)+\bar F(\bar z)~.
\ee

Differentiating  \eqref{s,t map} in $\tau, \,\s$ and using \eqref{F^-=}, we obtain
\be\ba\label{dz=}
&\dot z=\f{\rho^+(\a\bar F'^{\,2}-\bar \rho^{\,+\,2})}{\a\left[(\rho^+\,\bar F')^2-(\bar \rho^{\,+}\, F')^2\right]}~,\qquad
&\acute z=\f{\rho^+(\a\bar F'^{\,2}+\bar \rho^{\,+\,2})}{\a\left[(\rho^+\,\bar F')^2-(\bar \rho^{\,+}\, F')^2\right]}~,\\[1mm]
&\dot {\bar z}=-\f{\bar \rho^+(\a F'^{\,2}- \rho^{+\,2})}{\a\left[(\rho^+\,\bar F')^2-(\bar \rho^{\,+}\, F')^2\right]}~,
&\acute{\bar z}=-\f{\bar \rho^+(\a F'^{\,2}+\rho^{+\,2})}{\a\left[(\rho^+\,\bar F')^2-(\bar \rho^{\,+}\, F')^2\right]}~.
\ea\ee
A similar differentiation of \eqref{q=F+bar F}, with the help of \eqref{dz=}, gives
\be\label{phi_d and phi_p} 
 \dot \phi=\f{\a\,\bar F'\,F'+\bar \rho^+\rho^+}{\a\left(\rho^+\,\bar F'+\bar \rho^+\,F'\right)}~, \qquad \acute{\phi}
 =\f{\a\,\bar F'\,F'-\bar \rho^+\rho^+}{\a\left(\rho^+\,\bar F'+\bar \rho^+\,F'\right)}~,
\ee
and they lead to 
\be\label{L^2=}
1+\a\acute \phi^{\,2}-\a\dot \phi^2=\f{\left(\rho^+\,\bar F'-\bar \rho^{\,+}\,F'\right)^2}{\left(\rho^+\,\bar F'+\bar \rho^{\,+}\,F'\right)^2}~.
\ee
The left hand side here defines the determinant of the induced worldsheet metric in static gauge 
and for regular surfaces it has to be positive.
Thus, for regular surfaces, the expressions $\rho^+\,\bar F'\pm\bar \rho^{\,+}\,F'$ 
 have no zeros. Note that these expressions have the same sign for a sufficiently large zero mode $p$. Assuming this, we get
\be\label{L=} 
 \sqrt{1+\a\acute \phi^{\,2}-\a\dot \phi^2}=\f{\rho^+\,\bar F'-\bar \rho^{\,+}\,F'}{\rho^+\,\bar F'+\bar \rho^{\,+}\,F'}~.
\ee
From \eqref{p and dot q} then follows
\be\label{Pi=} 
 \Pi =\f{\a\,\bar F'(\bar z)\,F'(z)
 +\bar \rho^+\rho^+}{\a\left[\rho^+\,\bar F'(\bar z)-\bar \rho^+\,F'(z)\right]}~,
\ee
and using \eqref{dz=} we obtain
\be\label{Pi pm phi_p} 
\f{1}{2}\left( \acute{\phi}+\Pi \right)=\acute{z}\,F'(z)~,
\qquad \f{1}{2}\left( \acute{\phi}-\Pi \right)=\acute{\bar z}\,\bar{F}'(\bar z)~.
\ee

Equation \eqref{s,t map}, for a fixed $\tau$, defines $z$ and $\bar z$ as functions of $\s$.
For example, when the non-zero modes of $F'$ and $\bar{F}'$ are not excited, 
\be\label{inverse map}
z=\f{\tau}{\sqrt{1+\a\,p^2}}+\s~, \qquad \bar{z}=\f{\tau}{\sqrt{1+\a\,p^2}}-\s~.
\ee
In general, writing these functions as $z=\z(\s)$, $\bar{z}=\bar{\z}(-\s)$, we find that
they are  monotonic $\z'(x)>0$, $\bar{\z}'(\bar x)>0$ and obey the monodromies
\be\label{z(s+2pi)=}
\z(x+2\pi)=\z(x)+2\pi~, \qquad \bar\z(\bar x+2\pi)=\bar\z(\bar x)+2\pi~,
\ee
related to diffeomorphisms of a circle.
In the next subsection we show that \eqref{Pi pm phi_p} realizes a time dependent canonical map between the two gauges. 

Concluding this subsection we express the energy-momentum density components 
in the static gauge \eqref{P_m in static gauge} in terms of the light-cone gauge variables, using 
\eqref{dz=}, \eqref{phi_d and phi_p} and \eqref{L=}. With \eqref{P_m=cal P},
${\cal P}^2$ is obtained from \eqref{Pi=} and 
\bea\label{P_m in light-cone}\nonumber
&&{\cal P}^0=\f{(\a\bar F'^{\,2}+\bar \rho^{\,+\,2})(\a F'^{\,2}+ \rho^{\,+\,2})}{\a^2\left[(\rho^+\,\bar F')^2-(\bar\rho^{\,+}\, F')^2\right]}
=\acute{z} \left(\f{\rho^+}{\a}+\f{F'^{\,2}}{\rho^+}\right)=-\acute{\bar z} \left(\f{\bar\rho^{\,+}}{\a}+\f{\bar F'^{\,2}}{\bar\rho^{\,+}}\right),
\\[1mm]
&&{\cal P}^1=-\f{(\a\bar F'\,F'+\bar \rho^{\,+}\rho^+)(\a\bar F'\,F'-\bar\rho^{\,+}\rho^+)}{\a^2\left[(\rho^+\,\bar F')^2
-(\bar\rho^{\,+}\, F')^2\right]}
\\ \nonumber
&&\hspace{5cm}=\acute{z}\left(\f{\rho^+}{\a}-\f{F'^{\,2}}{\rho^+}\right)-\f{1}{\a}
=-\acute{\bar z} \left(\f{\bar\rho^{\,+}}{\a}-\f{\bar F'^{\,2}}{\bar\rho^{\,+}}\right)+\f{1}{\a}\,.
\eea
We will use these relations in the next section to relate the static and light-cone gauges in the Hamiltonian formulation.

\subsection{Hamiltonian approach to the compactified 3d string}

We now consider the Hamiltonian treatment of the same system. 
In the first order formulation of 3d string dynamics the action is
\be\label{String action-1}
S=\int \mbox{d}\tau\int_0^{2\pi}\f{\mbox{d}\s}{2\pi}\left[{\cal P}_\mu\,\dot X^\mu
-\l_1\,{\cal C}_1 - \l_2\,{\cal C}_2\right]~,
\ee
where $\l_1$, $\l_2$ are Lagrange multipliers and ${\cal C}_1$, ${\cal C}_2$ are the Virasoro constraints
\be\ba\label{Virasoro constraints}
&{\cal C}_1=({\cal P}\,\acute{X})~,\qquad
&{\cal C}_2=\f{1}{2}
\left[\a^2({\cal P}\,{\cal P})+(\acute X\, \acute X)\right].
\ea\ee
The compact coordinate $X^1$ has the expansion (for $L=1$)
\be\label{Expansion of X^2}
X^1=\s+\sum_{n\in \zz}q_n\,e^{-\text{i}\,n\,\s}~,
\ee
with $q_{-n}=q_n^*$, while the canonical momenta ${\cal P}_\m$ and the coordinates ($X^0$, $X^2$) remain periodic.  
They have the standard  mode expansion without the $\s$ term in \eqref{Expansion of X^2}.

It follows from the canonical Poisson brackets on the extended phase space
\be\label{canonical PB}
\{{\cal P}_\m(\s_1), X^\n(\s_2)\}=2\pi\,\d_\m^{~\n}\,\d(\s_1-\s_2)~,
\ee
that the Poisson brackets of the constraints \eqref{Virasoro constraints} form the algebra \eqref{P,H-PB}
\be\ba\label{PB C_1,2}
&\{{\cal C}_1(\s_1), {\cal C}_1(\s_2)\}=2\pi\big[{\cal C}_1(\s_1)+{\cal C}_1(\s_2)\big]\d'(\s_1-\s_2)\,, \\[1mm]
&\{{\cal C}_1(\s_1), {\cal C}_2(\s)\}=2\pi\big[{\cal C}_2(\s_1)+{\cal C}_2(\s_2)\big]\d'(\s_1-\s_2)\,,\\[1mm]
&\{{\cal C}_2(\s_1), {\cal C}_2(\s_2)\}=2\pi\,\a^2
\big[{\cal C}_1(\s_1)+{\cal C}_1(\s_2)\big]\d'(\s_1-\s_2)\,,
\ea\ee
and one  has to complete these first class constraints by gauge fixing conditions 
in order to eliminate non-physical degrees of freedom.

This can be done by  the Faddeev-Jackiw reduction in static gauge $X^0=\tau,\, X^1=\s$. 
For this one computes 
${\cal P}_\mu\,\dot X^\mu$
on the constrained surface ${\cal C}_1={\cal C}_2=0$ in this gauge.
The action \eqref{String action-1} then reduces to
\be\label{Reduced Str-action}
S|_{\text{st.g.}}=\int \mbox{d}\tau\int_0^{2\pi}\f{\mbox{d}\s}{2\pi}\left({\cal P}_0
+{\cal P}_2\,\dot{X}^2 \right),
\ee
where ${\cal P}_0$ becomes a function of the reduced canonical variables 
$({\cal P}_2 ,{X}^2)$.
Hence, ${\cal P}^0=-{\cal P}_0$ plays the role of the Hamiltonian density.

In order to relate the reduced Hamiltonian system to the deformed model, we rescale the canonical variables,
\be\label{X^1-paramaterization}
\,{\cal P}_2=\f{\Pi}{\sqrt{\a}}~, \qquad X^2=\sqrt{\a}\,\phi~,
\ee
and rewrite the constraints \eqref{Virasoro constraints} as
\be\label{Reduced constraints}
{\cal C}_1=\Pi\,\phi'+{\cal P}_1=0~,\qquad
2\,{\cal C}_2=\a(\Pi^2+\acute\phi^2)+\a^2{\cal P}_1^2-\a^2{\cal P}_0^2+1=0~.
\ee
These equations define the remaining phase space variables 
\be\label{P_0=}
{\cal P}_1=-\Pi\,\phi'~, \quad  {\cal P}_0=-\f{1}{\a}\,\sqrt{1+\a\left(\Pi^2+\acute\phi^2\right)+\a^2\left(\Pi\,\acute{\phi}\right)^2}~
\ee
and we finally obtain 
\be\label{Reduced Str-action=}
S|_{\text{st.g.}}=\int \mbox{d}\tau\int_0^{2\pi}\f{\mbox{d}\s}{2\pi}\left[
\Pi\,\dot{\phi}-\left({\cal H}_\a+\f{1}{\a}\right)\right]\,.
\ee
${\cal H}_\a$ is the Hamiltonian density of the deformed model \eqref{D-FF H}.
Thus, the Faddeev-Jackiw reduction of the compactified 3d string in the static gauge  leads to the deformed free-field model. 

We now consider Hamiltonian reduction of \eqref{String action-1} in light-cone gauge.
Introducing the light-cone coordinates 
\be\label{light-cone}
X^\pm=X^0\pm X^1~, \qquad {\cal P}_\pm=\f{1}{2}\left({\cal P}_0 \pm {\cal P}_1\right)~,
\ee
the string action \eqref{String action-1} and the constraints become
\bea\label{String action-2}
S=\int \mbox{d}\tau\int_0^{2\pi}\f{\mbox{d}\s}{2\pi}\left[{\cal P}_+\,\dot X^++{\cal P}_{-}\,\dot X^{-}+{\cal P}_2\,\dot{X}^2
-\l_1\,{\cal C}_1 - \l_2\,{\cal C}_2\right]~,\quad \\
\noalign{\noindent with}
\label{Virasoro constraints 2}
{\cal C}_1={\cal P}_+\,\acute{X}^+ + {\cal P}_-\,\acute{X}^-+{\cal P}_2\,\acute{X}^2,
\quad
{\cal C}_2=\f{1}{2}
\left[\a^2{\cal P}_2^2+\acute{X}_2^2-4\,\a^2\,{\cal P}_+{\cal P}_- -\acute{X}^+\,\acute{X}^-
\right].
\eea
Using the gauge freedom, we can eliminate the non-zero modes of ${\cal P}_-(\s)$ and $X^+(\s)$, 
similarly to the uncompactified case. Taking into account that $X^1$ has winding number one, the light-cone 
gauge condition reads
\be\label{l-c gauge 1}
X^+(\s)=-2\,\a\,{\cal P}_-\tau+\s~, \qquad \acute{\cal P}_-(\s)=0~.
\ee
This provides $\acute X^+(\s)=1$ and ${\cal P}_-(\s)=p_-$, where $p_-$ is the zero mode of 
${\cal P}_-(\s)$. Rescaling then the canonical variables similarly to \eqref{X^1-paramaterization}\footnote{Note that the 
pairs $(\Pi, \phi)$ and $(\varPi, \varPhi)$ differ from each other, though they denote the same variables 
in the initial extended phase space.}
\be\label{X^1=varphi}
\,{\cal P}_2=\f{\varPi}{\sqrt{\a}}~, \qquad X^2=\sqrt{\a}\,\varPhi~,
\ee 
the constraints \eqref{Virasoro constraints 2} can be written as
\be\label{Virasoro constraints 3}
{\cal C}_1={\cal P}_+ +  p_-\,\acute{X}^-+{\cal P}=0,
\quad
2\,{\cal C}_2=
2\,\a\,{\cal H}-4\,\a^2\,p_-\,{\cal P}_+ -\acute{X}^-
=0~,
\ee 
with
\be\label{H,P=}
{\cal P}=\varPi\acute{\varPhi}~, \qquad {\cal H}=\f 1 2\left(\varPi^2+\acute\varPhi^2 \right)~.
\ee
By \eqref{Virasoro constraints 3} one finds ${\cal P}_+$ and $\acute{X}^-$ in terms of $(\varPi,\varPhi)$ and the zero mode $p_-$
\be\label{P_+=}
{\cal P}_+=-\f{2\,\a\, p_-{\cal H}+{\cal P}}
{1-4\,\a^2 p_-^2}~, \qquad
\acute{X}^-=\f{2\,\a\, {\cal H}+4\,\a^2 p_- {\cal P}}{1-4\,\a^2 p_-^2}~.
\ee
The zero modes of the constraints \eqref{Virasoro constraints 3} satisfy
\be\label{zero mode eq}
(p_+-p_- )+ P=0~, \qquad 2\,\a\,H+(1-4\,\a^2\,p_-\,p_+)=0~,
\ee
where  $p_+$ is the zero mode of ${\cal P}_+$ and
\be\label{H,P}
P=\int_0^{2\pi} \f{\text{d}\s}{2\pi}\, {\cal P}~, \qquad H=\int_0^{2\pi}\f{\text{d}\s}{2\pi}\,{\cal H}~.
\ee
The string energy then becomes
\be\label{Str-energy}
E_{\text{str}}=-(p_+ + p_-)=\f{1}{\a}\,\sqrt{1+2\,\a\, H+\a^2 P^2}~.
\ee
Faddeev-Jackiw reduction of the action \eqref{String action-2} by the constraints 
\eqref{Virasoro constraints 2}-\eqref{l-c gauge 1} yields
\be\label{l-c reduced action}
S|_{\text{l-c.\,g.}}=\int \mbox{d}\tau\int_0^{2\pi}\f{\mbox{d}\s}{2\pi}\left[\varPi(\s)\,\dot{\varPhi}(\s)
-2\,\a\, p_+(p_-+\dot{p}_-\tau) 
+p_-\dot{x}^-\right],
\ee
where $x^-$ is the zero mode of the periodic part of $X^-(\s)$, and we have used the rescaled variables \eqref{X^1=varphi}.  
Neglecting the total derivative term $\f{\rm{d}}{\rm{d}{\tau}}(-2\,\a\, p_+ p_- \tau)$ in \eqref{l-c reduced action}, 
we obtain 
\be\label{l-c reduced action=}
S|_{\text{l-c.\,g.}}=\int \mbox{d}\tau\int_0^{2\pi}\f{\mbox{d}\s}{2\pi}\left[\varPi(\s)\,\dot\varPhi(\s)
+p_-\dot q^--2\,\a\, p_+\,p_-\right].
\ee 
with $q^-=x^-+2\,\a\,p_+ p_-\tau$.
Using \eqref{zero mode eq} and neglecting also the constant term $1/(2 \a)$,
we end up with the action 
\be\label{FF-action}
S|_{\text{l-c.\,g.}}=\int \mbox{d}\tau\int_0^{2\pi}\f{\mbox{d}\s}{2\pi}\left[\varPi\,\dot\varPhi
+p_-\dot q^- -{\cal H}\right]~,
\ee
where ${\cal H}$ is the free-field Hamiltonian density \eqref{H,P=} and $p_-$ is obtained from \eqref{zero mode eq}
\be\label{p_m}
p_{-}=\f{1}{2}\left(P-\f{1}{{\a}}\,\sqrt{1+2\,\a\, H+\a^2 P^2}\right)~.
\ee

The situation here is similar to the uncompactified case, where instead  of \eqref{p_m} one has the level matching condition 
$P=h-\bar h=0$.\footnote{In the Hamiltonian formulation the light-cone gauge is not a complete gauge fixing for the 
closed string. The constraint corresponding to the remaining gauge freedom is the level matching condition.   
After complete gauge fixing one arrives at a conformal gauge and the Hamiltonian 
formulation is then equivalent to the Lagrangian formulation in light cone gauge \cite{Thorn,Gleb}.}  
Further Hamiltonian reduction in both cases is inconvenient. One has to 
quantize the free-field model together with the 
particle $(p_-, q^-)$ and impose
the condition \eqref{p_m} at the quantum level. Note that the right hand side in \eqref{p_m} is a well defined operator 
in the Fock space of the free-field theory.

We now discuss the relation between the static and light-cone gauges in the Hamiltonian approach.
In general, reduced Hamiltonian systems obtained in two different gauges are related to each other by a canonical 
transformation generated by the constraints of the initial gauge invariant system.
Our aim is to describe the canonical map between the light-cone and the static gauges of the compactified 3d string.  

First note that the Virasoro constraints \eqref{Virasoro constraints} can be represented in the form 
\be\label{constrainnts}
{\cal C}(\s):=f_\m(\s)\,f^\m(\s)=0~, \qquad  \bar{\cal C}(\s):=\bar f_\m(\s)\,\bar f^\m(\s)=0~,
\ee
with
\be\label{f-bar f}
f^\mu(\s)=\f{1}{2{\sqrt{\a}}}\left(\a\,{\cal P}^\m(\s)+\acute{X}^\m(\s)\right)\,, \quad
\bar{f}^\mu(\s)=\f{1}{2{\sqrt{\a}}}\left(\a{\cal P}^\m(-\s)-\acute{X}^\m(-\s)\right)\,. 
\ee  
From the canonical Poisson brackets \eqref{canonical PB} follows 
\bea\nonumber
&&\{{\cal C}(\s_1), f^\m(\s)\}=2\pi\,\p_\s[f^\m(\s)\,\d(\s_1-\s)],\quad 
\{\bar{\cal C}(\s_1), \bar f^\m(\s)\}=2\pi\,\p_\s[\bar f^\m(\s)\,\d(\s_1-\s)],\\[1mm]
\label{generators of GT}
&&\{{\cal C}(\s_1), \bar f^\m(\s)\}=\{\bar{\cal C}(\s_1), f^\m(\s)\}=0.
\eea
The corresponding infinitesimal transformations 
\be\label{inf map}
f^\m(\s)\mapsto f^\m(\s)+\p_\s\left[\e(\s)\,f^\m(\s)\right]~, \qquad \bar f^\m(\s)\mapsto \bar f^\m(\s)+
\p_\s\left[\bar\e(\s)\,\bar f^\m(\s)\right]~,
\ee
lead to the global ones
\be\label{gauge tr}
f^\m(\s)\mapsto \z'(\s)\, f^\m(\z(\s))~, \qquad  \bar f^\m(\s)\mapsto \bar\z'(\s)\, \bar f^\m(\bar\z(\s))~,
\ee
where $\z(\s),~\bar\z(\s)$ are diffeomorphisms of the unit circle. 
Note that, in general, the group parameters $\e(\s),~\bar\e(\s)$ could be functions on the phase space, 
since the transformations are on-shell.

The static gauge provides the following parameterization of $f^\m$ and $\bar f^\m$
\be\ba\label{static gauge f}
&f^\m_{\text{st}.g.}(\s)=\f{1}{2}\left(\begin{array}{c}\sqrt{\a}\,{\cal P}^{0}(\s)
\\[1mm]  \sqrt{\a}\,\,{\cal P}^1(\s)+\f{1}{\sqrt{\a}} \\[1mm]  
\Pi(\s)+\phi'(\s)
\end{array}\right),\quad
\bar f^\m_{\text{st}.g.}(\s)=\f{1}{2}\left(\begin{array}{c}\sqrt{\a}\,{\cal P}^{0}(-\s)
\\[1mm] \sqrt{\a}\,{\cal P}^1(-\s)-\f{1}{\sqrt{\a}} \\[1mm] 
\Pi(-\s)-\phi'(-\s) \end{array}\right),
\ea\ee
where ${\cal P}^{0}$ and ${\cal P}^1$ are given by \eqref{P_0=}.

The light-cone gauge parameterization of $f^\m$ and $\bar f^\m$ is obtained from \eqref{l-c gauge 1}-\eqref{P_+=}
\be\ba\label{L-C gauge of f}
&f^\m_{\text{l-c.g.}}(\s)=\f{1}{2}\left(\begin{array}{c} \f{\rho^+}{\sqrt{\a}} +\f{\sqrt{\a}[\varPi(\s)+\varPhi'(\s)]^2}{4\rho^+}
\\[1mm] \f{\rho^+}{\sqrt{\a}} -\f{\sqrt{\a}[\varPi(\s)+\varPhi'(\s)]^2}{4\rho^+}\\[1mm]
\varPi(\s) +\varPhi'(\s)
\end{array}\right),
&\bar f^\m_{\text{l-c}.g.}(\s)=\f{1}{2}\left(\begin{array}{c} \f{\bar\rho^{\,+}}{\sqrt{\a}} 
+\f{\sqrt{\a}[\varPi(-\s)-\varPhi'(-\s)]^2}{4\bar\rho^{\,+}}
\\[1mm]
\f{\bar\rho^{\,+}}{\sqrt{\a}} -\f{\sqrt{\a}[\varPi(-\s)-\varPhi'(-\s)]^2}{4\bar\rho^{\,+}}\\[1mm]
\varPi(-\s) -\varPhi'(-\s)
\end{array}\right),
\ea\ee
where we have used
\be\label{rho}
2\rho^+=1-2\,\a\,p_-~, \qquad  2\bar\rho^{\,+}=-1-2\,\a\,p_-~.
\ee
Based on \eqref{gauge tr}, we introduce the relations
\be\label{gauge tr equations}
f^\m_{\text{st.g.}}(\s)=\z'(\s)\, f^\m_{\text{l-c.g.}}(\z(\s))~, \qquad 
\bar f^\m_{\text{st.g.}}(\s)=\bar \z'(\s)\, f^\m_{\text{l-c.g.}}(\bar\z(\s))~,
\ee
which by \eqref{static gauge f}-\eqref{L-C gauge of f} are equivalent to
\be\ba\label{zeta'}
&\a\left[{\cal P}^0(\s)+{\cal P}^1(\s)\right]+1=2\rho^+\z'(\s)\,,\\[1mm]
&\a\left[{\cal P}^0(\s)-{\cal P}^1(\s)\right]-1=\f{\a}{2\rho^+}\,\z'(\s)[\varPi(\z(\s))+\acute{\varPhi}(\z(\s))]^2\,,\\
&\Pi(\s)+\acute{\phi}(\s)=\z'(s)\left(\varPi(\z(\s))+\acute{\varPhi}(\z(\s))\right)\,,
\ea\ee
and similarly for the anti-chiral part
\be\ba\label{bar zeta'}
&\a\left[{\cal P}^0(-\s)+{\cal P}^1(-\s)\right]-1=2\bar\rho^{\,+}\bar\z'(\s)\,,\\[1mm]
&\a\left[{\cal P}^0(-\s)-{\cal P}^1(-\s)\right]+1=\f{\a}{2\bar\rho^{\,+}}\,\bar{\z}'(\s)[\varPi(-\bar\z(\s))+\acute{\varPhi}(-\bar\z(\s))]^2\,,\\[1mm]
&\Pi(-\s)+\acute{\phi}(-\s)=\bar\z'(\s)\left(\varPi(-\bar\z(\s))+\acute{\varPhi}(-\bar\z(\s))\right)\,.
\ea\ee
The integration in \eqref{zeta'} over $\s$ provides the relations
\be\label{E, P eq}
\a\left(E_{\text{str}}+P^1_{\text{str}}\right)+1=2\rho^+\,,\quad
\a\left(E_{\text{str}}-P^1_{\text{str}}\right)-1=\f{2\a h}{\rho^+}\,,
\ee
which for the string energy leads again to \eqref{deformed energy}. The same result is obtained for the antichiral part \eqref{bar zeta'}.

Equations \eqref{zeta'}-\eqref{bar zeta'} are equivalent to \eqref{Pi pm phi_p} and \eqref{P_m in light-cone} 
with $z(\s)=\z(\s)$ and $\bar{z}(\s)=\bar\z(-\s)$,
which indicates that they define a canonical map between the two gauges. The direct computation with the help 
of \eqref{s,t map}-\eqref{q=F+bar F} shows that this map preserves the canonical symplectic form 
\be\ba\label{canonical map}
&\int_{0}^{2\pi}\f{\text{d}\s}{2\pi}\,\d\Pi(\s)\wedge\d\phi(\s)=\int_{0}^{2\pi}\f{\text{d}\s}{2\pi}\,\d\varPi(\s)\wedge\d\varPhi(\s)=\\[1mm]
&\int_{0}^{2\pi}\f{\text{d}x}{2\pi}\,\d F(x)\wedge\d F'(x)+
\int_{0}^{2\pi}\f{\text{d}\bar x}{2\pi}\,\d \bar F(\bar x)\wedge\d \bar F'(\bar x)+
\f{1}{2}\,\d p\wedge \left[\d F(0)+\d\bar F(0)\right]\,.
\ea\ee

\section{Generalization to 2d CFTs and to (non-conformal) models with a potential}

In this section we first generalize the scheme described in Section 4.2  to other 2d CFTs.
Recall that starting from the free field model we had arrived at the $T\bar T$ deformed action. This was identified 
with the Nambu-Goto action of a 3d string in static gauge. We then wrote the unfixed 
NG action in Hamiltonian form and fixed the light-cone gauge. 
Faddeev-Jackiw reduction of the gauge fixed action lead to 
the original free field Hamiltonian. 

Guided by this, starting from a 2d CFT with a canonical description, specified by a 
Hamiltonian density ${\cal H}(\Pi,\phi,\acute\phi)$,
we will devise a first order system such that after going to static gauge we recover the deformed 
Hamiltonian while when working in light-cone gauge we arrive at the undeformed Hamiltonian 
${\cal H}$.  We then apply the same 
scheme to the model \eqref{H with U(q)} with a potential, which explicitly breaks conformal symmetry.
Relevant references for this section are \cite{Frolov1,Frolov2,Sfondrini2}. 

\subsection{Integrability of the deformed 2d CFTs}

We introduce a constrained Hamiltonian system with a string type action
\be\label{S-ext}
S=\int \mbox{d}\tau\int_0^{2\pi}\f{\mbox{d}\s}{2\pi}\left[{\cal P}_0\,\dot X^0+{\cal P}_{1}\,\dot X^{1}
+\Pi_k\,\dot\phi^k-\l_1\,{\cal C}_1 - \l_2\,{\cal C}_2\right]~,
\ee
where
\be\ba\label{Constraints}
&{\cal C}_1={\cal P}_0\,\acute{X}^0+{\cal P}_{1}\,\acute{X}^{1}+ {\cal P} ~,
\\[1mm]
&{\cal C}_2=\f{1}{2}
\left[\a^2\left({\cal P}_{1}^2-{\cal P}_0^2\right)+\left(\acute X_{1}^2-\acute X_0^2\right)\right]+\a\,{\cal H}(\Pi,\phi,\acute{\phi})~.
\ea\ee
${\cal H}$ and ${\cal P}$ are the Hamiltonian and momentum densities of a 2d CFT.  
We assume that the conditions \eqref{symmetry condition}-\eqref{traceless condition} are fulfilled.
Because of \eqref{P,H-PB} the Poisson brackets of the constraints \eqref{Constraints} satisfy 
\eqref{PB C_1,2}. 

The system is reparametrization invariant (with the appropriate transformation 
properties of $\lambda_{1,2}$ \cite{MW}). 
This enables us to introduce the static gauge,
where again $X^{1}$ is a compact coordinate. 
Doing this and applying the Faddeev-Jackiw reduction
one finds that the action \eqref{S-ext} reduces 
to the $T\bar T$-deformed system \eqref{Def-Action_H} with
the Hamiltonian density ${\cal H}_\a+{1}/{(2\a)}$, where ${\cal H}_\a$ as in eq. \eqref{H_a}.

If we fix instead light-cone gauge \eqref{l-c gauge 1}
and use the definitions \eqref{light-cone}, 
we arrive again at \eqref{FF-action}.  
The equations \eqref{Virasoro constraints 3}-\eqref{FF-action} are trivially generalized with the replacements
\be\label{replacements}
\varPi\,\dot\varPhi \mapsto \varPi_k\,\dot\varPhi^k~, \qquad \ \varPi\,\acute\varPhi \mapsto \varPi_k\,\acute\varPhi^k~, 
\qquad \f 1 2\left(\varPi^2+\acute\varPhi^2\right)\mapsto {\cal H}(\varPi,\varPhi,\acute{\varPhi})~.
\ee

\subsection{Generalization to models with a potential}

We now generalize the above discussion to theories with a conformal symmetry breaking potential $U(\phi)$. 
To this end we introduce a string like dynamical system such that in static gauge it
reduces to the deformed theory specified by the Hamiltonian density \eqref{H_a with U(q)}.

Consider an action of the type \eqref{S-ext}, where the constraint ${\cal C}_1$ is the same as in
\eqref{Constraints} but with a modified ${\cal C}_2$ of the form
\be\label{C_2}
{\cal C}_2=\f{1}{2}
\Big[g({\cal P}_{1}^2-{\cal P}_0^2)+\f{1-g^2b^2}{g}(\acute X_{1}^2-\acute X_0^2)-2\,b\,g({\cal P}_0 \acute X^{1}+
{\cal P}_{1} \acute X^{0})+
2\,{\cal H}\Big]\,.
\ee
This has the structure of the Hamiltonian density 
\eqref{Sigma model H} which guarantees that the constraints ${\cal C}_1$ and ${\cal C}_2$ 
satisfy the algebra \eqref{PB C_1,2}.
The matrices $G$ and $B$ in the space spanned by  ($X^0, X^{1}$), are
\be\label{G,B matrices}
G^{kl}=g\left(\begin{array}{cr} -1&0\\ \phantom{-}0 & 1
\end{array}\right)~, \qquad
B_{kl}=b\left(\begin{array}{cr} \phantom{-}0&1\\ -1& 0
\end{array}\right).
\ee
As before the system is reparametrization invariant and we can fix either static or light-cone gauge. 

The Faddeev-Jackiw reduction in the static gauge is again straightforward. 
If we identify \cite{Frolov1,Sfondrini2}
\be\label{g,b=}
g=\b~, \qquad b=\f{\a\, U(\phi)}{2\,\b} ~,\qquad \b=\a\left(1-{\a\over 2}U\right)
\ee
it leads to the Hamiltonian system \eqref{Action_H}
with the deformed Hamiltonian \eqref{H_a with U(q)}. 

We now turn to the reduction in light-cone gauge. 
The precise form of this gauge choice is less obvious in the non-conformal case and to find it we 
rewrite the first order system in second order Lagrangian form as a sigma-model with target space
coordinates $(X^0,X^1,\phi^k)$:
\be\label{Laction}
S=S[\phi]+{1\over 2\pi}\int d\tau d\s\Big(-{1\over 2\b(\phi)}\p_z X^+\p_{\bar z} X^-
+{1\over\a}\big(\dot X^0\,\acute X^1-\acute X^0\dot X^1\big)\Big) 
\ee 
where the first term is the 2d CFT action and the 
last term does not contribute to the equations of motion. 
For the light-cone fields $X^\pm$ they are 
\be
\p_z\left({1\over\b(\phi)}\p_{\bar z}X^-\right)=0\,,\qquad
\p_{\bar z}\left({1\over\beta(\phi)}\p_z X^+\right)=0\,,
\ee 
which can be integrated once 
\be
{1\over\b(\phi)}\p_{\bar z}X^-=\rho^-(\bar z)\,,\qquad
{1\over\beta(\phi)}\p_z X^+=\rho^+(z)\,.
\ee
$\rho^+$ and $\rho^-$ transform as one-forms under reparametrizations of the circle. 
Assuming that they have constant sign, which poses a restriction on the potential,  
one can gauge away the non-constant (oscillator) parts. 
In light-cone gauge $\rho^\pm$ are (arbitrary) constants. 

If we insert this into the equation of motion for $\phi$, we obtain 
\be\label{Liouville}
{\delta\over\delta\phi^k}S[\phi] +{1\over 4}\a^2\rho^+\!\rho^-{\p\over\p\phi^k} U(\phi)=0\,.
\ee
For appropriate choice for $\rho^\pm$ these are the equations of motion of the undeformed 
theory. 
In the case of a single scalar field $\phi$ with a free action and potential   
\be
U(\phi)=2-2\,e^{2\,\phi}
\ee
equation \eqref{Liouville} becomes the Liouville equation. 

For the same choice of potential and $\a=1$, the action \eqref{Laction} (before gauge fixing) 
and ignoring the boundary term is the $SL(2)$ WZW-model \cite{Wipf,Marc}.

\bigskip

\noindent
{\bf Acknowledgements} 

\medskip

\noindent
We would like to thank Harald Dorn and Alessandro Sfondrini for useful discussions. 
G.J. thanks MPI for Gravitational Physics in Potsdam for warm hospitality during his visits
in 2018 and 2019 and S.T. is grateful to the Mathematical Institute of TSU for supporting a visit 
in Tbilisi.  
\appendix

\setcounter{equation}{0} 
\def\theequation{A.\arabic{equation}}

\section{Solution for the light-cone chiral fields}

Due to \eqref{modes of F}, the Fourier mode expansions of $F'^{\,2}(z)$ and $\bar F'^{\,2}(\bar z)$ 
\be\label{FF L_n}
F'^{\,2}(z)=\sum_{n\in\zz} L_n\,\text{e}^{-\mathrm{i}nz}~, \qquad\bar F'^{\,2}(\bar z)
=\sum_{n\in\zz}\bar L_n\,\text{e}^{-\mathrm{i}n\bar z}~,
\ee
define $L_n$ and $\bar L_n$ as the Virasoro generators in the standard free-field form
\be\label{FF L_n=}
L_n=\f{1}{2}\sum_{n\in\zz}a_m\,a_{n-m}~, \qquad \bar L_n=\f{1}{2}\sum_{n\in\zz}\bar a_m\,\bar a_{n-m}~,
\ee
with $a_0=\bar a_0={p}$. The solution of \eqref{F^-=} can then be written as
\be\label{Phi^-=}
\Phi^-(z)=\rho^-z+\f{\mathrm{i}\a}{\rho^+}\sum_{n\neq 0}\f{L_n}{n}\,\text{e}^{-\mathrm{i}nz}~,\qquad
\bar\Phi^{\,-}(\bar z)=
\bar\rho^{\,-}z+\f{\mathrm{i}\a}{\bar\rho^+}\sum_{n\neq 0}\f{\bar L_n}{n}\,\text{e}^{-\mathrm{i}n\bar z}~,
\ee
where $\rho^-$ and $\bar\rho^{\,-}$ are given by \eqref{rho^-=}. 
We neglect the constant zero modes of $\Phi^-(z)$ and $\bar\Phi^{\,-}(\bar z)$;  
they correspond to translations of $X^0$ and $X^2$.

\setcounter{equation}{0}
\def\theequation{B.\arabic{equation}}

\section{String energy in the static and light-cone gauges}

The integration of \eqref{P_m in light-cone} over $\s$, for a fixed $\tau$,  yields
\be\ba\label{Str-E}
\f{1}{\a}\int_0^{2\pi} \f{\text{d}\s}{2\pi}\, \f{1+\a\acute \phi^{\,2}}{\sqrt{1+\a\acute \phi^{\,2}-\a\dot \phi^2}}
=\f{1}{\a}\int_0^{2\pi}\f{\text{d}z}{2\pi} \left(\rho^+ +\f{\a F'^{\,2}(z)}{\rho^+}\right)=\f{1}{\a}(\rho^+ +\rho^-)
~~~\\[1mm]
~~~~=\f{1}{\a}\int_0^{2\pi}\f{\text{d}\bar z}{2\pi} \left(\bar \rho^{\,+} 
+\f{\a \bar F'^{\,2}(\bar z)}{\bar \rho^{\,+}}\right)=\f{1}{\a}(\bar \rho^{\,+} +\bar\rho^{\,-}).
\ea\ee
According to \eqref{string energy},  the left hand side of this equation is the string energy 
in the static gauge
and the right hand sides correspond to the string energy in the light-cone gauge
\eqref{string energy n=1}. This straightforward calculation confirms the validity 
of \eqref{deformed energy}, without referring to the gauge invariance of the string energy.

A similar calculation for the string momentum $P^1$ by \eqref{P_m in light-cone} yields
\be\ba\label{Str-P^1}
P^1=\int_0^{2\pi} \f{\text{d}\s}{2\pi}\, \f{-\dot{\phi}\,\acute\phi}{\sqrt{1+\a\acute \phi^{\,2}-\a\dot \phi^2}}
=\f{1}{\a}\int_0^{2\pi}\f{\text{d}z}{2\pi} \left(\rho^+ -\f{\a F'^{\,2}(z)}{\rho^+}\right)=
\\[1mm]
\f{1}{\a}(\rho^+ -\rho^--1)=\bar{h} -h.
\ea\ee




\end{document}